\documentclass[aps,prb,twocolumn,showpacs]{revtex4-1}
\usepackage{graphics}
\usepackage{subfigure}
\usepackage{epsfig}
\usepackage{epsf,epic}
\usepackage{color}
\usepackage{marvosym }
\usepackage{subfigure}
\usepackage{amsmath}
\usepackage{amssymb}
\usepackage{amsfonts}
\usepackage{wrapfig}
\usepackage{multirow}
\usepackage{bm}
\usepackage{dcolumn}
\newcommand{\etal}{\textit{et al.\ }}
\newcommand{\ie}{\textit{i.e.\ }}

\begin{document}
\title{Lattice polarization effects on the screened Coulomb interaction $W$ of 
the GW approximation}
\author{Walter R. L. Lambrecht}
\affiliation{Department of Physics, Case Western Reserve University, 10900 Euclid Avenue, Cleveland, OH-44106-7079}
\author{Churna Bhandari}
\affiliation{Department of Physics, University of Missouri,
Columbia, Missouri 65211, USA}
\author{Mark van Schilfgaarde}
\affiliation{Department of Physics, King's College London, London WC2R 2LS, United Kingdom}
\begin{abstract}
In polar insulators where longitudinal and transverse optical phonon modes
differ substantially, the electron-phonon coupling affects the energy-band
structure primarily through the long-range Fr\"ohlich contribution to the Fan
term.  This diagram has the same structure as the $GW$ self-energy where $W$
originates from the electron part of the screened coulomb interaction. The two
can be conveniently combined by combining electron and lattice contributions
to the polarizability.  Both contributions are nonanalytic at the origin,
and diverge as $1/q^2$ so that the predominant contribution comes from a small
region around ${\bf q}=0$.  Here we adopt a simple estimate for the Fr\"ohlich
contribution by assuming that the entire phonon part can be attributed to a
small volume of ${\bf q}$ near ${\bf q}=0$.  We estimate the magnitude for
$\mathbf{q}{\rightarrow}0$ from a generalized Lyddane-Sachs-Teller relation, and
the radius from the inverse of the polaron length scale.  The gap correction is
shown to agree with Fr\"ohlich's simple estimate $-\alpha_P\omega_\mathrm{LO}/2$ of the polaron
effect with $\alpha_P$ the polaron couping factor. 
\end{abstract}
 \maketitle
\section{Introduction}
The $GW$ approximation\cite{Hedin65,Hedin69} provides one of
the most successful many-body-perturbation theoretical approaches to the
electronic band structure of solids. It is based on an expansion of the
self-energy in the screened Coulomb interaction $W$.  In fact, the self-energy
schematically is approximated as $\Sigma=iGW$ with $G$ the one-electron Green's
function.  In its most recent quasiparticle self-consistent all-electron
version,\cite{vanSchilfgaarde06,Kotani07} which we label QS$GW$, it has become applicable to a wide
arrange of systems.  Nonetheless, it significantly overestimates the band gap in
 strongly ionic systems.  This effect has been attributed to the neglect of
ladder diagrams, which attract electron-hole pairs, enhance the screening, which
reduces $W$ and the splitting between occupied and unoccupied states (see
e.g. Ref.~\onlinecite{Shishkin07}).  However, the lattice polarization also
contributes to $W$ and enhances the screening. Previous estimates of
both of these effects in literature \cite{Shishkin07,PasquarelloChen15,Botti13}
attribute most of the band gap overestimate by straight $GW$ to just one of these
but did not consider them together. 

Both electronic and polaronic
terms are bosonic in origin, have the same diagrammatic structure and can be
conveniently combined into a single $GW$ diagram where $W$ includes both
electronic and lattice contributions to the polarizability.  This fact was first
used by Bechstedt\cite{Bechstedt05} to modify $W$ when solving both the
$GW$ band structure and the Bethe-Salpeter
equation for the polarizability in strongly ionic materials. 
They apply the effect only in the static limit ($\omega=0$) and within the
Coulomb-hole static screened exchange (COHSEX)
framework. Furthermore using a model dielectric function, their correction
amounts to replacing the macroscopic $\varepsilon_\infty$ by $\varepsilon_0$
in the ${\bf q}\rightarrow0$ limit. 
Vidal \etal \cite{Vidal10} estimated the
renormalization of band gaps in materials with large Fr\"ohlich
coupling parameter.  They adopted Bechstedt's approach 
and estimate a gap reduction about 1 eV in CuAlO$_2$.  
Subsequently Botti and
Marques (BM) made a refinement, taking into account the dynamics in $W$
by using a generalized Lyddane-Sachs-Teller relation.\cite{Botti13} However,
they did not properly take into account the volume confinement of $W$ in ${\bf q}$-space. 

The main physics was already laid out by Hedin \cite{Hedin69} in the framework of many-body
perturbation theory. He partitioned out the Fan term as a separate contribution to the self-energy,
$iGW^\mathrm{ph}$; and indeed this has been the customary approach. The main effects on temperature
dependent band structure and the zero-point motion corrections have been worked out in the
Allen-Heine-Cardona (AHC) theory.\cite{AllenHeine76,AllenCardona81,AllenCardona83}  More recently
both Fan and Debye-Waller contributions have been implemented in a density-functional framework.
\cite{Antonius15,Marini15}  A recent review by Giustino \cite{Giustino17} describes the different
approaches to the electron-phonon coupling problem and points out the relations between the
adiabatic AHC theory and the more general Hedin\cite{Hedin69} and Baym \cite{Baym61} field
theoretical approaches and their modern implementation. The latter rely on interpolation of the
electron-phonon coupling coefficients on a fine {\bf k}-point integration mesh by means of maximally
localized Wannier functions.\cite{GiustinoCohenLouie} This approach however becomes problematic for
the long-range parts of the electron-phonon coupling in polar materials, the so-called Fr\"ohlich
part,\cite{Frohlich54,Vogl76,Verdi15,Sjakste15} because the latter decay as $1/q$ and lead to a
divergent $1/q^2$ contribution (and which becomes $1/q^4$ near band extrema if applied
straightforwardly to the AHC equations). While these problems can be overcome by removing the
adiabatic approximation,\cite{Giustino17,Ponce15jchemp} it seems worthwhile pursuing simpler
approaches to estimate the Fr\"ohlich part of the Fan term, which dominates in compounds where
$\varepsilon_\infty$ is small compared to $\varepsilon_\mathrm{tot}$.

Noting that the main problem with the Fr\"ohlich term occurs near band edges, 
Nery and Allen\cite{NeryAllen16} developed a ${\bf k}\cdot{\bf p}$ approach for the band
near the band-edge, dividing the Fan term into a non-analytic Fr\"ohlich part and a remainder.
Using the resulting
simple form of the Fr\"ohlich term, the singular integral can be done analytically and can then
be combined with the numerical integration without the need for an excessively fine integration mesh.
We here take their approach a step further toward simplification. As they showed, the crucial
length-scale for the effect is the polaron length-scale. Therefore, we use the inverse polaron length
directly as integration limit for the singular Fr\"ohlich term.  While this is a more approximate
estimate than their, in principle, exact approach, which subtracts the singularity from
the numerical integration and replaces it by the analytical result for a simple model, we use
directly the simple model and estimate the size of the region in ${\bf q}$-space where it is
applicable as the inverse polaron length. 

The BM approach is another option worth revisiting. This is the main goal of our paper. 
We identify the problem in BM's treatment of the ${\bf q}\rightarrow 0$ limit. Again,
our solution is to base this on the polaron-length scale.  We show that the BM approach, as is, 
depends crucially on the size of the integration mesh and the gap correction decreases
proportional to $1/N_{\rm mesh}$ and would thus go to zero at convergence. Instead, we assume
that the correction applies in a finite ${\bf q}$-region of size the inverse polaron length.
  The advantage compared to a full evaluation of
the Fr\"ohlich contribution to the Fan terms is that the computational
effort is vastly simpler; moreover the phonon contribution to the entire band structure is
obtained in a single calculation. 

We apply both approaches to a set of strongly ionic materials, MgO, NaCl, LiF, LiCl
and show that the modified BM approach  leads to results in good agreement with the above
simplified Nery-Allen polaronic estimate. We also consider zincblende GaN for a direct comparison to Nery
and Allen's more complete approach, although the effect here is an order of magnitude smaller. 
Admittedly, our approach does not address the
full electron-phonon coupling renormalization of the band gap, only the Fr\"ohlich part.
However, for strongly ionic materials, this is arguably the largest contribution. The other
electron-phonon contributions to the zero-point motion correction are large only in
systems with only light atoms. 
Finally,
we consider the relative importance of this effect to the effects of missing
electron-hole interactions based on literature data. The conclusion is that the latter
are in fact a more important correction to the band-gap reduction.

\section{Theory}
As is well known,\cite{Maradudin71,Gonze2} optical phonon modes can strongly
modify the screening in polar compounds.  This is nicely encapsulated by the
generalized Lyddane-Sachs-Teller (LST) relation in the $\mathbf{q}{\rightarrow}0$ limit
\begin{equation}
\frac{\varepsilon_{tot}^{\alpha}(\mathbf{q}\rightarrow 0,\omega)}{\varepsilon_{el}^{\alpha}(\mathbf{q}\rightarrow 0,\omega)}=\prod_m\frac{\left(\omega^\alpha_{\mathrm{LO};m}\right)^2-\omega^2}{\left(\omega_{\mathrm{TO};m}\right)^2-(\omega+i0^+)^2}. \label{eqLST} 
\end{equation}
The product runs over all optical modes $m$ which are infrared active and have a
longitudinal-transverse splitting
($\omega_{\mathrm{LO};m}>\omega_{\mathrm{TO};m}$) and belong to the irreducible
representation corresponding to the polarization direction $\alpha$.  The
superscript $\alpha$ indicates the direction along which $\mathbf{q}{\rightarrow}0$
approaches zero and the LO modes depend on this direction. We next examine, how this
affects the screened Coulomb interaction $W$ in the $GW$ theory.

In practical calculations, $W(\mathbf{q},\mathbf{r},\mathbf{r}',\omega)$ is represented by an expansion in
a basis set.  In our all-electron implementation\cite{Kotani07} this consists of
a mixed product basis with Bloch sums of products of of partial waves inside
augmentation spheres and plane waves in the interstitial.  Thus $W$ becomes a
matrix $W_{\mu\nu}(\mathbf{q},\omega)$.  More commonly plane waves are used for
these bosonic degrees of freedom, in which case $\mu$ becomes ${\bf G}$.
As noted already the effect is dominant in the $\mathbf{q}{\rightarrow}0$ limit.
Treatment of $W_{\mu\nu}(\mathbf{q}{\rightarrow}0,\omega)$ requires special care because
of the divergence of the Coulomb interaction $v(\mathbf{q})=4\pi e^2/q^2$.
(There is a similar divergence for Fr\"ohlich contribution.)  It is however integrable,
because what is needed for the self-energy is a convolution integral over both
energy and wave vector,
\begin{eqnarray}
\Sigma^c_{nm}({\bf k},\omega)&=&\frac{i}{2\pi}\int d\omega'\sum_\mathbf{q}^\mathrm{BZ}\sum_{n'}^\mathrm{all} 
G_{nn'}({\bf k}-\mathbf{q},\omega-\omega')\nonumber \\
&&\sum_{\mu\nu}
W^c_{\mu\nu}(\mathbf{q},\omega')e^{-i\delta\omega'} \nonumber \\
&&\langle\psi_{{\bf k}n}|\psi_{{\bf k}-\mathbf{q}n'}E_\mu^\mathbf{q}\rangle
\langle E_\nu^\mathbf{q}\psi_{{\bf k}-\mathbf{q}n'}|\psi_{{\bf k}m}\rangle.\label{eq:sig}
\end{eqnarray}
The sum over $\mathbf{q}$ becomes an integral 
and the contribution from the region near $\mathbf{q}=0$ over a small sphere 
multiplies  the divergence  by $q^2dq$.  Here the Green's function
$G_{nn'}({\bf k}-\mathbf{q},\omega-\omega')=[\omega-\omega'-\epsilon_{{\bf k}-\mathbf{q}}\pm i\delta]^{-1}\delta_{nn'}$, 
is a diagonal matrix in the basis of one-electron states,$\psi_{{\bf k}n}$;
the screened Coulomb interaction  is expanded in 
an auxiliary mixed product basis set 
$E_\mu$ which diagonalizes the bare Coulomb matrix,\cite{Friedrich10} 
and conversion factors from one basis to the other are included. 
The superscript $c$ refers to taking the correlation part $W^c=W-v$.  

The approach dealing with this integrable divergence has been described 
by Freysoldt \etal\cite{Freysoldt06} in the context of a 
plane wave basis set expansion of the bare and 
screened Coulomb interaction and by 
Friedrich \etal \cite{Friedrich09,Friedrich10} 
in terms of a mixed product basis set expansion.
The method consists in replacing the integral over the BZ by an exactly 
integrable function with the same type of divergence. The difference 
between the two is then a smooth function for which the integral can 
be replaced by a discrete sum.  The approach 
originally was introduced by Massida \etal\cite{Massida} in Hartree-Fock calculations
because, in fact, the same problem already affects the bare exchange.
In the context of $GW$ theory, it requires a knowledge of the 
screened Coulomb interaction near $\mathbf{q}=0$ and this can 
either be obtained by an analytical ${\bf k}\cdot{\bf p}$ approach\cite{Friedrich10} or 
by using the offset-$\Gamma$ method. 
The actual approach
used in the QS\emph{GW} program\cite{ecalj} is described in 
Kotani \etal\cite{Kotanijpsj} and provides an improved version 
of the offset-$\Gamma$ method used in Ref. \onlinecite{Kotani07}.
To obtain the behavior $W(\mathbf{q})$ near $\mathbf{q}=0$, one needs 
the macroscopic inverse dielectric constant, which is 
 $1/\varepsilon^{-1}_{00}(\mathbf{q}{\rightarrow}0,\omega)$. It is calculated
by a block matrix inversion separating out the divergent term from 
the known behavior of the polarizability matrix as function of $\mathbf{q}$. 
Here the subscripts  $00$ of the 
dielectric function matrix refer to the reciprocal lattice vector 
${\bf G}=0$ in a plane wave basis set, or equivalently the first mixed-product
basis set function in the basis set that diagonalizes the bare Coulomb 
interaction.\cite{Friedrich10}  Both of these in fact correspond to the 
average over the unit cell. 
The inverse of this quantity is then expanded in spherical harmonics
and only the $L=00$ spherical average is required for the integral of 
the ``head'' of $\varepsilon^{-1}_{00}$. This is if we neglect some 
higher order corrections, discussed by Betzinger \etal\cite{Betzinger10}

For a simple estimate of the phonon contribution, we use the fact that (1)
the Fr\"ohlich contribution to $W$ originates predominately from the divergent,
small-$q$ region,\cite{NeryAllen16} and (2) we handle this region using the
usual formulation of the $GW$ self-energy calculation through a special
treatment of the integrable divergence of $W$ in the neighborhood of
$\mathbf{q}=0$ only.  Eq.~(\ref{eq:sig}) is integrated numerically on a discrete
$\mathbf{q}$ mesh, and the ``central'' cell term is treated specially to handle
the divergence.  Our approach simply modifies the central cell dielectric
function $\varepsilon^{-1}_{00}(\mathbf{q}{=}0)$ using the appropriate Lyddane-Sachs-Teller
factor.  The fact that its limit is non-analytic, \ie depends on the direction
of $\mathbf{q}$ means that it is a second rank tensor with non-zero components
dictated by symmetry.  For example for an orthorhombic crystal, it will have
only diagonal components but the $xx$, $yy$ and $zz$ diagonal elements are all
different.  In general in the anisotropic offset-$\Gamma$
method\cite{Kotanijpsj} it is expanded in invariant tensors corresponding to the
symmetry of the cell and requires at most six $\mathbf{q}$ points close to
$\mathbf{q}=0$ where the macroscopic dielectric constant must be evaluated and
for which we need to know the corresponding LO-TO splittings.

In a full approach to the electron-phonon coupling, one can arbitrarily cut out
some small region near the singularity, subtract the standard mesh integration
technique result for that region and replace it by the properly integrated
singularity. This is the approach followed by Nery and Allen.\cite{NeryAllen16}
Here we focus exclusively on the Fr\"ohlich term, and thus we cannot rely on a
cancellation of the two treatments to the self-energy integral. We thus need an
accurate estimate of the range of the polaron effect.  In the treatment of the
$\mathbf{q}{\rightarrow}0$ limit for the purely electronic screening, the
relative weight of the specially treated $\Gamma$-cell depends on the size of
the $\mathbf{q}$-point mesh. The finer the mesh, the less the weight of the
$\Gamma$-cell.  The electron-phonon contribution should not depend on the mesh
spacing, but since we lump the entire contribution into the central cell and
omit contributions from other microcells, the Fr\"ohlich contribution to
$W_{00}(\mathbf{q}{=}0,\omega)$ should be rescaled as described below.

We may decouple the convergence in $\mathbf{q}$-space of the electronic polarizability from that of
the phonon contribution as follows. We define the LST factor $f_{\rm LST}$ to be the factor that
corrects $W$, so that the additive correction is $\Delta W=(f_{\rm LST}-1)W$.  At $\mathbf{q}{=0}$,
$f_{\rm LST}$ is the inverse of the factor in Eq. (\ref{eqLST}). Now, according to the above
discussion, we want to correct the $\mathbf{q}{=0}$ value of $W$ but this represents an effective 
volume in \textbf{q}-space. So, we need to estimate separately the size of this ${\bf q}$-space region
over which the effect of the lattice polarization is to be taken into account.
Let's call this $q_\mathrm{LP}$ and the corresponding
volume $\Omega_\mathrm{LP}=q_\mathrm{LP}^3$. When we calculate the
convolution integral $\Delta\Sigma({\bf k})=\sum_{\mathbf{q}}G({\bf k}-\mathbf{q})\Delta
W(\mathbf{q})$ as a discrete sum, we assume only the $\mathbf{q}{=0}$ microcell of volume
$\Omega_{GW}=\Omega_{BZ}/N_\mathrm{mesh}$ contributes, so $\Delta\Sigma({\bf k})=iG({\bf k})\Delta W(0)$.
If the GW mesh is coarser than $q_{LP}$, than we might overestimate the
effect. On the other hand if it is finer, than the phonon correction should be extended to GW-mesh
points beyond ${\bf k}=0$. Instead, we may simply rescale $\Delta{W}(0)$ to
$W(0)\Omega_\mathrm{LP}/\Omega_{GW}$.  This means that for the pure electronic screening part, the usual
compensation between the discrete sum (non-divergent part) and the special treatment of the
$\Gamma$-cell is still valid. But for the added $\Delta\Sigma=i(f_{\rm LST}-1)G({\bf k})W(0)\Omega_\mathrm{LP}/\Omega_{GW}$ we use a fixed volume of $\mathbf{q}$-space corresponding to $q_\mathrm{LP}$

The essential problem to obtain meaningful results within this approach is thus to pick $q_\mathrm{LP}$. We note that it cannot be obtained from
considerations of the phonons or of the dependence of $\varepsilon_{tot}({\bf q})/\varepsilon_{el}({\bf q})$ alone because the latter lack information on
the electron-phonon coupling to the bands, which must involve the
effective masses of the bands. 
Following Nery and Allen's idea,\cite{NeryAllen16} the relevant length scale  here is the polaron length
$a_P=\sqrt{\hbar/2\omega_{LO} m_*}$, where $\omega_{LO}$ is the longitudinal phonon
for which we consider the electron-phonon contribution and $m_*$ is the band effective mass. 
This means there is actually a different polaron length for electrons and holes,
which we denote $a_{Pe}$ and $a_{Ph}$. 
Using an electron (hole) effective  mass of 0.35 (1.26)
for MgO as an example, we obtain polaron length scales of 20.84 and 10.96 $a_0$ ($a_0$ is the Bohr radius).
We use $(2m_{hh}+m_{lh})/3$ along the [100] direction for the holes. We use this type of average
because the heavy hole band is doubly degenerate. 
The average $a_P=(a_{Pe}+a_{Ph})/2$
defines an inverse length scale of 0.06 $a_0^{-1}$ which for MgO is about 1/12 of the
BZ.  Typically, for a two atom unit cell system like MgO, a $8\times8\times8$ mesh already gives both the $GW$
and phonons very well converged. We also found
that the gap correction in the BM approach varies as $1/N_\mathrm{mesh}$.
We can thus extrapolate to the appropriate $q=1/a_P$ or use
the approach for decoupling from the mesh-size described in the
previous paragraph. The precise way of averaging the effective masses here is not crucial because
we only are  trying to estimate the polaron length and the results are not very sensitive to this
estimate.

The approach described here is similar to that of Botti and Marques \cite{Botti13}, except that they
did not take into account the range of the Fr\"ohlich interaction. In their formulation the effect
would have vanished in the limit of small microcell size.

They also appeared to confuse the relevant length scale: they said ``It is easy to understand that
the coupling of phonon waves and electromagnetic waves is effective only for ${\bf q}\rightarrow 0$,
since the speed of sound is negligible if compared with the speed of light.'' This is true but
rather irrelevant to the problem considered here.  In fact, the coupling of electromagnetic waves to
phonon waves, \ie polariton formation, following Pick\cite{Pick70}, occurs only when $q=|{\bf
q}|\gg\omega_0/c$ with $\omega_0$ a typical phonon frequency and $c$ the speed of light. Decoupling
occurs for $q\ll\omega_0/c$.  This is about $q=10^{-4}$ of the Brillouin zone (BZ). In other
words, this theory provides a cut-off {\em above} rather than below which the effect comes into
play.  However, we are not concerned with the retardation effects of polariton formation here, we
are interested in the applicability of the LST relation and the polaronic effect on the band gap. 

We may also directly estimate the Fr\"ohlich singularity integral.
Following Nery and Allen, \cite{NeryAllen16}
the singular contribution near the band edge to the zero-point motion self-energy is given by
\begin{equation}
  \Delta E_{n{\bf k}}=-\alpha_P \hbar \omega_\mathrm{LO} \tan^{-1}{(q_Fa_\mathrm{LO})}\frac{2}{\pi}, \label{eq:nery}
\end{equation}
where
\begin{equation}
  \alpha_P=\frac{e^2}{2a_P} \frac{1}{\hbar\omega_\mathrm{LO}}\left[\frac{1}{\varepsilon_\infty}-\frac{1}{\varepsilon_0}\right],
\end{equation}
is the dimensionless polaron coupling constant. The question now is what to use as integration
cut-off $q_F$ for the upper limit of the Fr\"ohlich singularity integral.  Clearly an overestimte 
will be obtained if we use $q_F=q_{BZ}=2\pi/a$ because  the expression is supposed to be valid only over the
region where the band dispersion is parabolic. It is even customary to let $q_F\rightarrow \infty$, in
which case we obtain $-\alpha\hbar\omega_\mathrm{LO}$ as upper limit.\cite{Frohlich54}
In Nery and Allen's approach this choice is not
crucial because they look at where this explicit approach becomes equivalent to the standard
integration approach of the other electron-phonon coupling terms besides the long-range Fr\"ohlich
one. This occurs at about 1/6 of the BZ in their case of GaN.  A better estimate would be
$q_F=1/a_P$; then the inverse tangent factor is simply $\pi/4$ and Eq.~(\ref{eq:nery})
simplifies to
\begin{eqnarray}
  \Delta E_{n{\bf k}}&=&-\alpha_P \omega_\mathrm{LO}/2,\nonumber \\
  &=&\frac{e^2}{4a_P}\left(\varepsilon_\infty^{-1}-\varepsilon_0^{-1}\right), \label{eqeshift}
\end{eqnarray}
where if $n$ corresponds to the conduction band minimum (CBM), we use $a_{Pe}$ and if $n$ corresponds to the
valence band maximum (VBM), we would use $a_{Ph}$ as polaron lengths. So, this is simply half the difference in
Coulomb interaction at the polaron length calculated with purely electronic screening and
electronic plus lattice screening. 
This differs from
the upper limit by only a factor of 2, so the estimation of
$q_F$ is not very crucial if our goal is to obtain the right order of magnitude.
For the example of GaN used by Nery and
Allen, we find this already gives an excellent approximation to their full calculation.
We will show that it also agrees well with the modified BM approach described above,
in which $q_{LP}$ is also set to $1/a_P$.

We emphasize again, that if one
applies the electron-phonon coupling fully at all \textbf{q}-points then one can subtract out the
region of the singularity from the mesh sum and add it back in integrated analytically. However, in
the BM approach we focus entirely on the singular contribution, and thus we are limited by how
reliably we estimate the size of the singularity. This is true both if one thinks of it as a
Fr\"ohlich coupling strength dipole singularity or as the $1/q^2$ singularity in the $W$ screened
Coulomb potential. In fact, both are essentially the same. and both give a correction proportional to
$\varepsilon_\infty^{-1}-\varepsilon_0^{-1}$ and inversely proportional to $a_P$.

As far as the frequency dependence is concerned, it is clear from Eq.(\ref{eqLST}) that for
$\omega>\omega_\mathrm{LO}$ the factor quickly goes to 1.  Therefore in the integral over frequency,
it is sufficient to apply the effect only at $\omega=0$ as long as the first non-zero $\omega$-mesh
point is already well above $\omega_\mathrm{LO}$.  If one wishes to apply the effect including its
frequency dependence, then one needs to use a sufficiently fine integration mesh near the
$\omega_\mathrm{TO}$ and $\omega_\mathrm{LO}$ phonons to carry out the integrals over these poles
correctly.  We have done both and find that reliable results can be obtained using a coarse mesh,
and scaling $W(\mathbf{q}{=0},\omega{=}0)$ only. This is further
discussed in the Appendix \ref{appC}. Since a very fine
frequency mesh is required if the pole is properly summed over, this greatly simplifies the
computational effort.

\section{Computational details}
All calculations before are carried out using the QS$GW$ approach in the LMTO basis set
implementation,\cite{Kotani07} which can be found on-line at Ref.\onlinecite{ecalj}.  The relevant phonons are
taken from experiment or can be calculated using the ABINIT program.\cite{abinit}

The experimental lattice constants were used, 
MgO (4.21 \AA\cite{Fei99}), NaCl (5.64 \AA\cite{Gray63}), LiCl (5.14 \AA\cite{Ewald}),
LiF (4.02 \AA\cite{Ewald}).
In the QS$GW$ calculation, for MgO we used semicore Mg $2p$ and high lying $3s$ states as local orbital for the completeness of basis set.
For NaCl, we also used semicore $2p$ as  local orbitals.

\section{Results and discussion}

\begin{table}
  \caption{Parameters used to calculate the polaron length and polaron coupling strength
    in various materials and the polaron shifts of the band edges and gap.
    Effective masses in units of free electron mass, $m_h=(2m_{hh}+m_{lh})/3$.
    Phonon frequency $\omega_\mathrm{LO}$ in cm$^{-1}$, polaron lengths $a_{Pe}$ and $a_{Ph}$
    in Bohr units $a_0$.
    Polaron coupling constants $\alpha_{Pe}$ and $\alpha_{Ph}$ for electrons and holes
    are dimensionless. Energy shifts in meV.
    \label{tabpolaron}}
  \begin{ruledtabular}
    \begin{tabular}{lccccc}
      &MgO& NaCl&LiF&LiCl&GaN \\ \hline
      $m_e$ &0.35& 0.35& 0.61 & 0.40 & 0.18 \\
      $m_{hh}[100]$ & 1.70 & 2.10 & 2.83 & 1.06 & 1.70  \\
      $m_{lh}[100]$ & 0.40 & 0.55 & 1.10 & 0.56 & 0.50  \\
      $m_h$ &1.26 & 1.58 & 2.25 & 0.89 & 1.30 \\
      $\omega_\mathrm{LO}$ (cm$^{-1}$)& 722 & 265 & 656 & 382 & 730  \\
      $a_{Pe}$ ($a_0$) & 20.8 & 34.4 & 16.6 & 26.8 & 28.9 \\
      $a_{Ph}$ ($a_0$) & 11.0 & 16.2 & 8.6 & 17.9 & 10.7 \\
      $a_P$   ($a_0$) & 15.9 & 25.3 & 12.6 & 22.4 & 19.8 \\
      $\varepsilon_\infty$ & 3.0 & 2.3 & 1.95 & 2.8 & 5.6  \\
      $\varepsilon_0$     & 9.8 & 5.9 & 9.0 & 11.2 & 9.9 \\
      $\alpha_{Pe}$    &  1.7 & 3.2& 4.1 & 2.9 & 0.4 \\
      $\alpha_{Ph}$    &  3.2 & 6.9& 7.8 & 4.4 & 1.1 \\
      $\Delta E_{CBM}$ (meV) &  75  & 53 & 165 & 69 & 18 \\
      $\Delta E_{VBM}$ (meV)&  144  & 113 &317 & 103 & 49 \\
      $\Delta E_g$   (meV)   &  219  & 166 & 483 & 172 & 67 \\
    \end{tabular}
  \end{ruledtabular}
\end{table}

In Table \ref{tabpolaron} we show the polaron lengths and coupling strengths for electrons
and holes as well as the parameters entering them. The effective masses are obtained from
fitting the QS$GW$ calculations before adding the lattice polarization correction. The
dielectric constants are taken from experiment but could also be calculated in DFT
using for example the ABINIT program or using the electronic band structure within QS$GW$
for $\varepsilon_\infty$ and calculating $\varepsilon_0$ using the LST factor. 
Finally it shows the estimated band edge and gap shifts
using Eq. \ref{eqeshift}. 
We note that for GaN, this amounts to 67  meV, close to Nery and Allen's own estimate of 50 meV,
especially when considering that we expect an errorbar of order 10 meV in view of the various
approximations made. 
We may note that typically the shift is larger for the VBM because of the shorter polaron length
because of the larger hole mass.
For NaCl, we may compare our result with Fr\"ohlich's own estimate
of the conduction band shift.\cite{Frohlich54} He used an effective
mass $m_e=1$ and obtained 0.18 eV. Using $m_e=1$ we would obtain 0.09 eV
differing by a factor $2$ because in Fr\"ohlich's estimate the upper limit
of the singularity integral is replaced by $\infty$.

\begin{table}
  \caption{Band gaps in LDA, QSGW, QSGW+LPC in the BM approach with different k-meshes and
    extrapolated to $q_{LP}=1/a_P$, as well as the zero-point motion (ZPM) correction to the gap.
    All values in eV. The \% change in QSGW-LDA gap correction due to LPC is also given.\label{tabgaps}}
  \begin{ruledtabular}
    \begin{tabular}{lccccc}
                      & MgO & NaCl & LiF & LiCl & GaN \\ \hline
      LDA &            4.65 & 5.01 & 9.43 & 6.32 & 1.76 \\ 
      QSGW &           8.69 & 9.44 & 16.19 & 10.19 &  3.54 \\
      QSGW+LPC-BM-6 &  7.99 & 8.80 & 14.87 & 9.52  &  3.31 \\
      QSGW+LPC-BM-8 &  8.20 & 8.99 & 15.32 & 9.71  &  3.50 \\
      QSGW+LPC &       8.43 & 9.26 & 15.64 & 9.98  &  3.50 \\
      ZPM-Fr\"ohlich  &      -0.26 & -0.18& -0.55 & -0.21 & -0.04 \\
      ZPM-polaron     &      -0.22 & -0.17 &-0.48 & -0.17 & -0.07 \\
      QSGW-LDA        & 4.04 & 4.43 & 6.76 & 3.87 & 1.78 \\
      \% change LPC & -6     & -4 & -8 & -5 & -2  \\ \hline
      QSGW-BM \footnote{From Botti and Marques \cite{Botti13}}  & 8.94 & 9.52 & 15.81 & 10.28& \\
      QSGW+LPC-BM$^a$                                           & 7.71 & 8.37 & 13.69 & 9.05 & \\
      ZPM-BM$^a$  &-1.23 & -1.15 & -2.12 & -1.23 & \\ \hline
      Expt.  gap  & 7.8 & 8.5 & 14.2 & 9.4 &  3.5$\pm0.1$ \\ 
    \end{tabular}
  \end{ruledtabular}
\end{table}

Next we compare the above polaron estimates of the band edge shift with the results of the
modified BM approach, in which we use the average polaron length $a_P$ to set the
$q_{LP}=1/a_P$.  In this table, we show the LDA gaps, the QS$GW$ gaps and the
QS$GW$ gaps with the BM-type lattice polarization correction using $N_\mathrm{mesh}=6$ or 8. We note that the QS$GW$ results are well converged already with
a $6\times6\times6$ mesh, which differ from the $8\times8\times8$ mesh
by only 0.1 eV. However, the LPC correction is then effectively applied only
to a smaller region and has less weight. Therefore the corresponding
QS$GW$+LPC-BM-8 gap is less reduced than the QS$GW$+LPC-BM-6 one.   
We then extrapolate from the $q_\mathrm{LP}=(2\pi/a)(1/N_\mathrm{mesh})$ to $q_\mathrm{LP}=1/a_P$
assuming linear dependence. This is the result labeled QSGW+LPC.
Finally, the zero-point-motion correction,  due to the
lattice-polarization correction, labeled ZPM-Fr\"ohlich in Table \ref{tabgaps}
is the difference between the QSGW+LPC and QSGW gaps
and should be compared with the polaron effect given in Table \ref{tabpolaron}.
Rounding the values of $\Delta E_g$ of Table \ref{tabpolaron} to 0.01 eV, a more realistic
estimate of the uncertainty, we obtain the results in the row labeled ZPM-polaron.  We can see that
the two estimates agree with each other to within a few 0.01 eV.
Comparing with the gaps and  gap reduction values given by Botti and Marques in the next few rows, 
we see that their calculation significantly overestimated the effect. This is primarily because
they used a $6\times6\times6$ mesh but there are also differences in the QS$GW$ results themselves
which result from their use of a pseudopotential approximation and a plane wave basis set compared
to our all-electron and LMTO basis set. Finally, we also give the experimental gaps.  

We note that our QS$GW$ gap
for MgO is lower than that of BM (8.94 eV) or Shishkin et al. \cite{Shishkin07}
(9.16 eV) or Chen and Pasquarello\cite{PasquarelloChen15} (9.29 eV). We note
that if we use a less converged basis set (leaving out the higher energy 3s
local orbitals on O for example, we also find a higher gap). We therefore caution that
comparisons of the lattice-polarization effect in $GW$ between different
methods should also keep in mind that differences
between all-electron and pseudopotential methods as well as various convergence issues may play a role.

Our adjusted gaps still overestimate the experimental gaps. This
indicates that the electron-hole effects on the gap reduction may be more
important than the lattice polarization correction. For MgO, the results of Shishkin \etal \cite{Shishkin07} and Chen \etal\cite{PasquarelloChen15}  indicate effects of the
order of 20\% of the QS\emph{GW}-LDA gap correction. The under-screening in the
QS$GW$ due to the lack of electron-hole interactions or random phase approximation, 
was noted before and for most semiconductors amounts to about 20\%.
This has led to a commonly used correction factor of
$0.8\Sigma$.\cite{Chantis06a,Deguchi16}.  A universal factor of 0.8 
can be approximately justified because $\varepsilon_{\infty}$ is uniformly underestimated
by a factor of 0.8 in a wide range of semiconductors.  The results of Chen \etal\cite{PasquarelloChen15}
and Shishkin\cite{Shishkin07} MgO and NaCl, and LiCl further support this.  Our
results indicate a further reduction of this gap correction by the lattice
polarization by about 5\%. The percentage reduction of the QSGW-LDA gap correction due to the
lattice polarization effect is given in Table \ref{tabgaps} and varies from 2-8 \%. 
Taken together this would reduce the
QS\emph{GW}-LDA gap correction by 25 \%. The zero-point motion
correction of the gap in MgO was previously estimated to be 0.15-0.19 eV for
MgO.\cite{Antonius15} but it is not clear whether this properly includes the
long-range Fr\"ohlich contribution. Assuming it is not and
adding the present Fr\"ohlich lattice polarization
correction, the total zero-point-motion correction would then amount to 0.4 eV.

\begin{figure}[!htp]
  \includegraphics[scale=0.275]{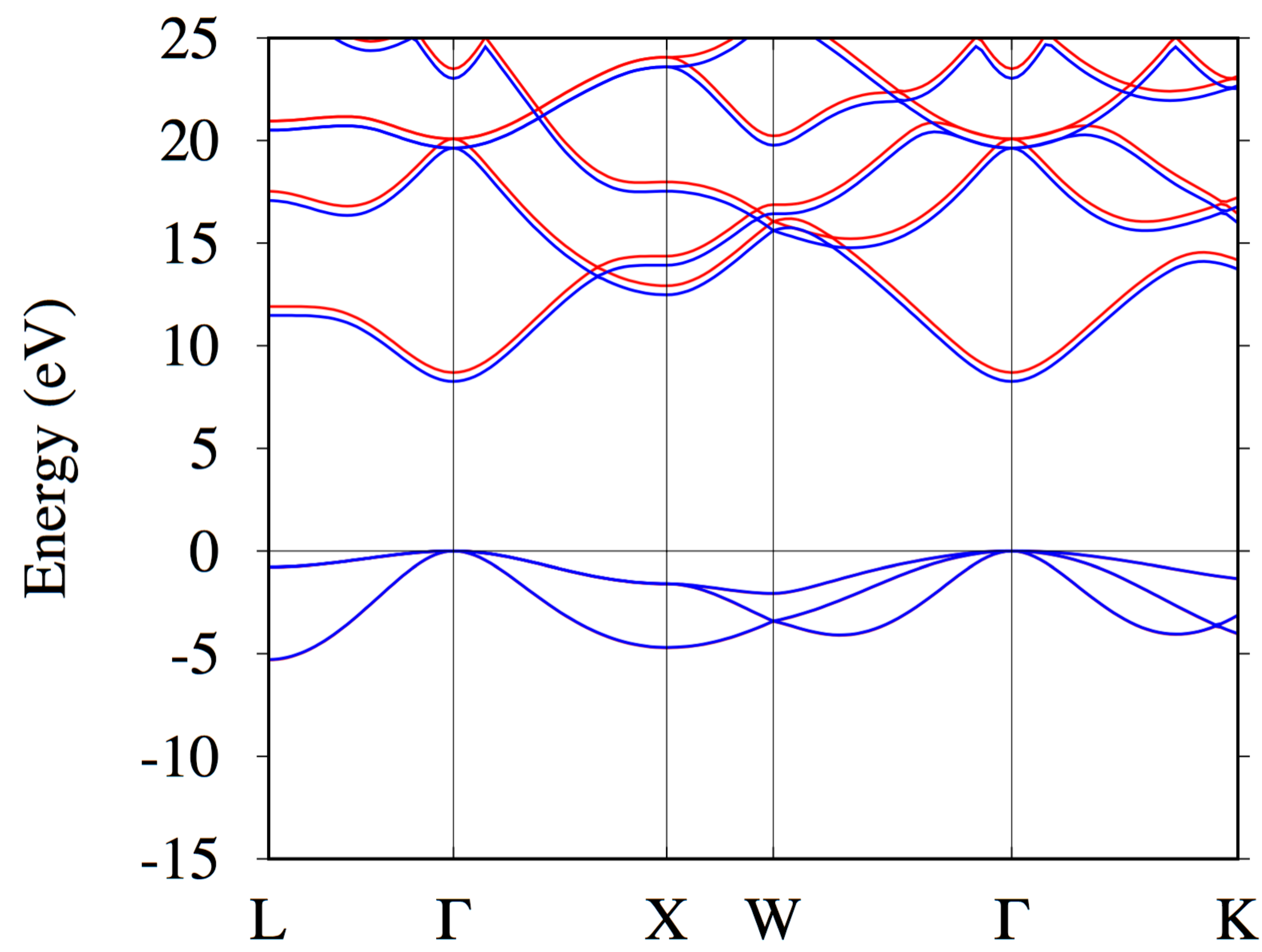} 
  \includegraphics[scale=0.275]{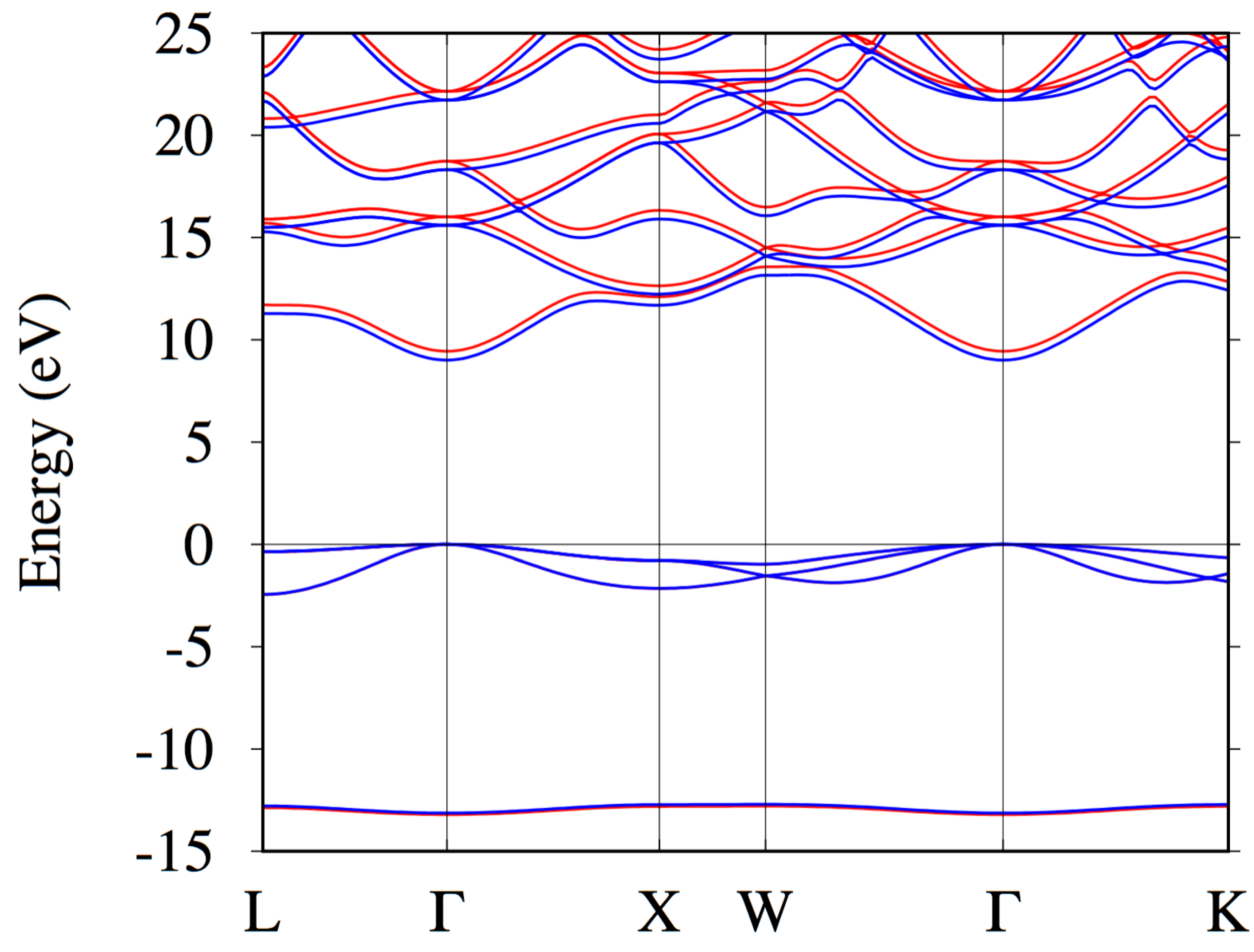}
 \caption{(Color-online) Band structure of MgO (top) and NaCl (bottom), the red color for QS$GW$ and blue for QS$GW$+LPC with $N_\mathrm{mesh}=8$.}\label{figband}
  \end{figure} 

One advantage of our modified BM approach is that in principle it not only corrects
the gap but the entire band structure.
Plotting the full band structures, as shown in Fig. \ref{figband} for two examples
MgO and NaCl, and for $N_\mathrm{mesh}=8$ we find that the effect amounts
pretty much to a constant shift of the conduction band once
the valence bands are aligned. Using the fixed mesh size may still somewhat overestimate
the effect on the gap as seen from Table \ref{tabgaps} but nonetheless allows us
to estimate how it affects the rest of the band structure. 
This constant shift is by no means guaranteed  because it is not clear
immediately whether the same polaron length-scale estimates of the required integration region
of the ${\bf q}\rightarrow0$ singularity, $q_\mathrm{LP}$ applies also to other bands.
Our calculation assumes that this can be taken the same for all bands. At present we are
not aware of experimental evidence to test this result of a constant shift. 

\section{Conclusions}
In this paper, we revisited the approach proposed by Botti and Marques\cite{Botti13} to
estimate the lattice polarization effect on $W$ in the $GW$ method and hence on the band structure
in ionic materials. As pointed out
by Giustino\cite{Giustino17} the BM approach is equivalent to the Fr\"ohlich contribution to the
Fan self-energy, which has thus far only received limited attention in spite of the large amount of
work on electron-phonon coupling renormalization of the band gaps of materials. This
is primarily due to the technical difficulties in calculating it, which require a very fine
${\bf k}$-space integration mesh. 

We develop here a simplied approach which takes advantage of the fact that the Fr\"olich interaction is dominant in a
small region around $\mathbf{q}$=0.  The effective volume of $\mathbf{q}$ is fixed by the polaron length scale
$a_P=\sqrt{\hbar/2\omega_\mathrm{LO} m_*}$.   With this length scale for the Fr\"olich interaction, and the LST
relations at $W(\mathbf{q}{=0})$, we can construct a simple description that modifies $W$ directly, 
which can be used both in $GW$ calculations and for higher order diagrams involving $W$,
e.g. incorporation of ladder diagrams via the Bethe-Salpeter equation \cite{Rohlfing98},
with minimal cost.

We compared our results of the BM method with a simple estimate based on directly integrating the Fr\"ohlich
electron-phonon coupling singularity near the band gap and find that the latter can also be estimated simply by setting
the cut-off of the singularity integral to the inverse polaron length.  We note that the present  method allows us to estimate
the gap correction to at best a few 0.01 eV only 
because of the remaining uncertainty in the polaron length $a_P$.
For a more refined treatment to meV prevision a full
electron-phonon calculation of the band shifts will be required
and would allow adjusting $a_P$ to fit it. 

\acknowledgments{This work was supported by the US Department of Energy, Office of Science, Basic
Energy Sciences under grant No.  DE-SC0008933.  Calculations made use of the High Performance
Computing Resource in the Core Facility for Advanced Research Computing at Case Western Reserve
University.  MvS was supported by EPSRC CCP9 Flagship Project No. EP/M011631/1.}

\appendix*

\section{Discussion of the $\omega\rightarrow0$ limit.} \label{appC}
Here, we address the issue of whether we need the $\omega$-dependence
of the LST factor.
As far as the frequency dependence is concerned, we can in principle include
the frequency dependent LST factor  as given in Eq.(1) in the main text.
However, in evaluating $\Sigma$ we then need to make 
sure the extra pole due to the phonon is properly integrated over. 
The method for calculating the energy integral in Eq.(2) in the main text
is described in detail in Sec. II.F of Ref. \onlinecite{Kotani07}. 
It is done with a contour integral mostly along the imaginary axis 
but including pole contributions from the energy bands along the real axis.
The inclusion of the lattice polarization effect through the LST factor
introduces poles in $W(\omega)$ close to $\omega=0$ in the energy 
range of the phonons, more precisely at the $\omega_{LO}$ frequencies. 
We thus need to ensure that these do not lead to spurious effects and are 
adequately integrated over. The behavior of $W$ near $\omega\rightarrow0$
along the imaginary axis is already taken care of specially in 
Ref. \onlinecite{Kotani07}. The band structure poles, however lead to 
a contribution $W^c(\omega-\epsilon_{{\bf k}-{\bf q}n})$. These are tabulated
on an $\omega$-mesh along the real axis for later interpolation of $\Sigma(\omega)$ to 
the values required. For example, in the QS$GW$, method one needs 
$\left[\Sigma_{nm}(\epsilon_{{\bf k}m})+\Sigma_{nm}(\epsilon_{{\bf k}n})\right]/2$.
Thus this mesh must be chosen fine enough so that adequate interpolation 
is possible if $\omega-\epsilon_{{\bf k}n}\approx0$ for some energy band. 
Unphysical values would result  if this energy band is close to the pole. 
One may avoid a divergence by adding a small  imaginary  part to the 
$\omega$ and using a fine mesh in the region of the phonons. However, 
for a reasonable $\omega$-mesh spacing for the electronic part of 
$W^c$, all except the first point $\omega=0$ are usually well above the 
phonon frequencies where the LST factor goes to 1. Thus, we may also only 
correct the $\omega=0$ mesh-point where the correction factor is simply 
$\prod_m\omega_{LOm}^2/\omega_{TOm}^2$.  We have tested both  approaches and 
found that they give the same result for the final band gap. 
In fact, intuitively, one does not expect the detailed behavior near 
each phonon to have a specific effect. Such an effect would occur 
whenever some energy band difference $\epsilon_{{\bf k}n}-\epsilon_{{\bf k}-{\bf q}n'}$ is close to an LO phonon energy and leads to an almost divergent 
contribution $W^c(\omega_{LO})$. This means that the $\omega$-dependence of the LST factor,
which is one of the distinguishing features of BM compared to the previous work of
Bechstedt\cite{Bechstedt05} who only applied the correction to the static part of $\Sigma$
is not as important as one might guess at first sight and Bechstedt's approach is adequate,
especially in view of the other approximations we are already making  and the overal goal
to keep this approach as simple as possible.

\begin{table}
\begin{ruledtabular}
\caption{The band gap (in eV) in MgO as function of $d\omega$ (in Rydberg), 
the mesh size along the real frequency axis used in the calculation
of $\Sigma$.}
\begin{tabular}{l|ccccc}
$d\omega$ &0.0002& 0.0005&0.004 &0.01 & 0.02  \\
$E_{g}$&8.06& 7.95 &7.89 &8.12 &8.08  \\
\end{tabular}
\label{tabmesh}
\end{ruledtabular}
\end{table}

We compare results with different real $d\omega$ mesh sizes for MgO. 
This $d\omega$ corresponds to the mesh spacing used in the interpolation 
of $\Sigma(\omega)$ as mentioned earlier. 
From Table \ref{tabmesh}, we see that for  
$d\omega=0.0002$ Ry the band gap is 8.06 eV which is very close to the value when we use a large mesh spacing 
  $d\omega=0.02$ Ry. So, either we pick the mesh so fine 
the phonon region is integrated and interpolated over correctly, or 
we pick it big enough so we just skip over it and effectively only 
include the correction at the first $\omega=0$ point. 
However, for intermediate values of $d\omega$, the gap seems 
to vary in a somewhat uncontrolled way. This results from the difficulty of 
interpolating the rapidly varying LST correction factor of 
Eq.(1) in the region near the phonon frequency. It is clear
that the latter goes through an asymptote at $\omega_T$, or its 
inverse goes through an asymptote at $\omega_L$. Thus the value 
can rapidly change from positive to negative and unreliable results 
are obtained if the integration mesh samples just a few random 
points on this curve. In fact, note that 
the $\omega_T\approx 0.003$ Ryd. in this case of MgO and hence 
$d\omega=0.0002$ is sufficiently fine compared to $\omega_T$ and 
$d\omega=0.02$ is sufficiently large to just skip over the whole range, 
while $d\omega\approx 0.004$ is troublesome.  We conclude from this 
that correcting only the $\omega=0$ value is more efficient and 
sufficiently accurate.  
Similar results are obtained for NaCl.  
However in that case, the relevant phonon frequencies are significantly 
smaller and hence an even finer and ultimately, unpractical 
mesh would be required.

\bibliography{lst,gw,lmto,abinit,srto}

\end{document}